\documentclass[a4paper,11pt]{article}
\usepackage{pos}
\RequirePackage{wrapfig}

\title{Various constraints on BSM physics from extensive air showers and from ultra-high energy gamma-ray and neutrino searches}
\ShortTitle{Constraints on BSM physics}

\author*[a]{Olivier Deligny}

\affiliation[a]{Laboratoire de Physique des 2 Infinis Ir\`ene Joliot-Curie (IJCLab)\\
CNRS/IN2P3, Universit\'{e} Paris-Saclay, Orsay, France}

\emailAdd{deligny@ijclab.in2p3.fr}

\abstract{Various phenomena of physics beyond that of the Standard Model could occur at high scale. Ultra-high energy cosmic rays are the only particles available to explore scales above a few dozens of TeV. Although these explorations are much more limited than those carried out with colliders, they provide a series of constraints in several topics such as tests of Lorentz invariance, dark matter, phase transitions in the early universe or sterile neutrinos. Several of these constraints are reviewed in these proceedings of UHECR2024 based on searches for anomalous characteristics in extensive air showers or searches for ultra-high energy gamma rays and neutrinos.}

\FullConference{7th International Symposium on Ultra High Energy Cosmic Rays (UHECR2024)\\
 17-21 November 2024\\
Malargüe, Mendoza, Argentina\\}

\begin{document}
\maketitle

\section{Introduction}

Despite its impressive successes, the Standard Model (SM) of particle physics is known to be incomplete. There are numerous reasons to expect Beyond-Standard-Model (BSM) physics to take place at high scale. Ultimately, ultra-high-energy cosmic rays (UHECRs) are the only laboratory for probing this physics. In these proceedings of UHECR2024, various investigations of high-scale BSM physics are reviewed. Rather than limiting ourselves to generic setups, the guiding thread is to rely on concrete models and to review the constraints placed by the absence of anomalous observations in UHECR physics on the fundamental parameters governing these models. 

Exact symmetries governing the SM Lagrangian guarantee invariance under Lorentz transformations and under the combination of charge conjugation, spatial inversion, and time reversal ($CPT$). The underlying symmetries are deeply connected in any quantum field theory~\cite{Luders:1954zz,Jost:1957zz,Schwinger:1958rc}. Lorentz invariance has always proved resilient to decades of tests remarkably accurate. Yet, theories that aim to describe Planck-scale physics ($\sim 10^{19}~$GeV) might break these fundamental symmetries~\cite{Kostelecky:1988zi}. Although the scales at which interactions take place in extensive air showers (EAS) are many orders of magnitude away from the Planck scale, small effects at low energy might still be observable. Constraints obtained from the data of the Pierre Auger Observatory are reviewed in Section~\ref{sec:LIV}.

UHECR interactions cause the production of pion mesons that subsequently decay into gamma rays and neutrinos, the search of which at energies around $10^{8}~$GeV is of primary importance to decipher further the origin of UHECRs. The production of gamma rays and/or neutrinos at even higher energies, $E \gtrsim 10^{10}~{\rm GeV}$, is expected from cascade processes triggered by, for instance, the decay of putative particles with superheavy masses, as reminded in Section~\ref{sec:UHEgnu}. Their detection may therefore be instrumental in uncovering BSM physics, such as superheavy dark matter, cf. Section~\ref{sec:SHDM}, or new gauge bosons and particles trapped in topological defects, cf. Section~\ref{sec:CS}.

BSM physics may also be revealed through heavy neutral leptons that would affect the rate of interactions of Earth-skimming neutrinos. The exploration of an astrophysical scenario that would provide an exceptional opportunity to detect $\sim~$deca-GeV sterile neutrinos is reviewed in Section~\ref{sec:up}.

Finally, some prospects on UHECR and BSM physics are addressed in Section~\ref{sec:future}.

\section{Searches for Lorentz invariance violation} \label{sec:LIV}

To assess and predict possible violations of Lorentz invariance, effective field theories such as the Standard Model Extension have been designed~\cite{Colladay:1998fq}. These models preserve gauge invariance to guarantee internal consistency such as non-negative probabilities but introduce Lorentz invariance violation through spontaneous symmetry breaking caused by scalar fields and/or through $CPT$ violation, leading to preferential reference frame effects. This could lead, for example, to changes in the dispersion relation, resulting in different maximum attainable velocities for different particles or in vacuum birefringence and parity violation. We present constraints on these effects obtained in a robust way from the analysis of EAS data. 

\subsection{Stable neutral pions?}

One emblematic implementation of Standard Model Extension is that of~\cite{Coleman:1998ti}, in which the Lagrangian invariant under $SU(3)\times SU(2)\times U(1)$ is constructed from all renormalisable terms that are furthermore rotationally and translationally invariant in a preferred frame identified with that in which an observer would observe no dipole-own motion in the cosmic microwave background. In this framework, the mismatch between low-energy mass eigenstates and high-energy-momentum ones translates into ``velocity-mixing'' effects that turn on gradually with energy, effectively giving rise to different maximum attainable velocities. In particular, the dispersion relations relating the energy $E_i$ to the momentum $\mathbf{p}_i$ for each particle $i$ are modified according to $E_i^2=m_i^2+(1+\eta_{i0}) \mathbf{p}_i^2$, with $m_i$ the mass of the particle and $\eta_{i0}$ a small coefficient that controls a Lorentz-violating term in the kinetic sector. Beyond this particular setup, and motivated by quantum gravity theories, $\eta$ can be thought as a Planck-suppressed expansion of the spatial derivative operator and the most general dispersion relations can be parameterized as~\cite{Aloisio:2000cm}\footnote{Natural units are used throughout this review.}
\begin{equation}
    E_i^2=m_i^2+|\mathbf{p}_i|^2+\sum_{n\geq 0} \eta_{in}\left(\frac{|\mathbf{p}_i|}{M_\mathrm{P}}\right)^n|\mathbf{p}_i|^2,
\end{equation}
with $M_\mathrm{P}$ the Planck mass. 

\begin{figure}[ht]
    \centering
    \includegraphics[width=0.7\linewidth]{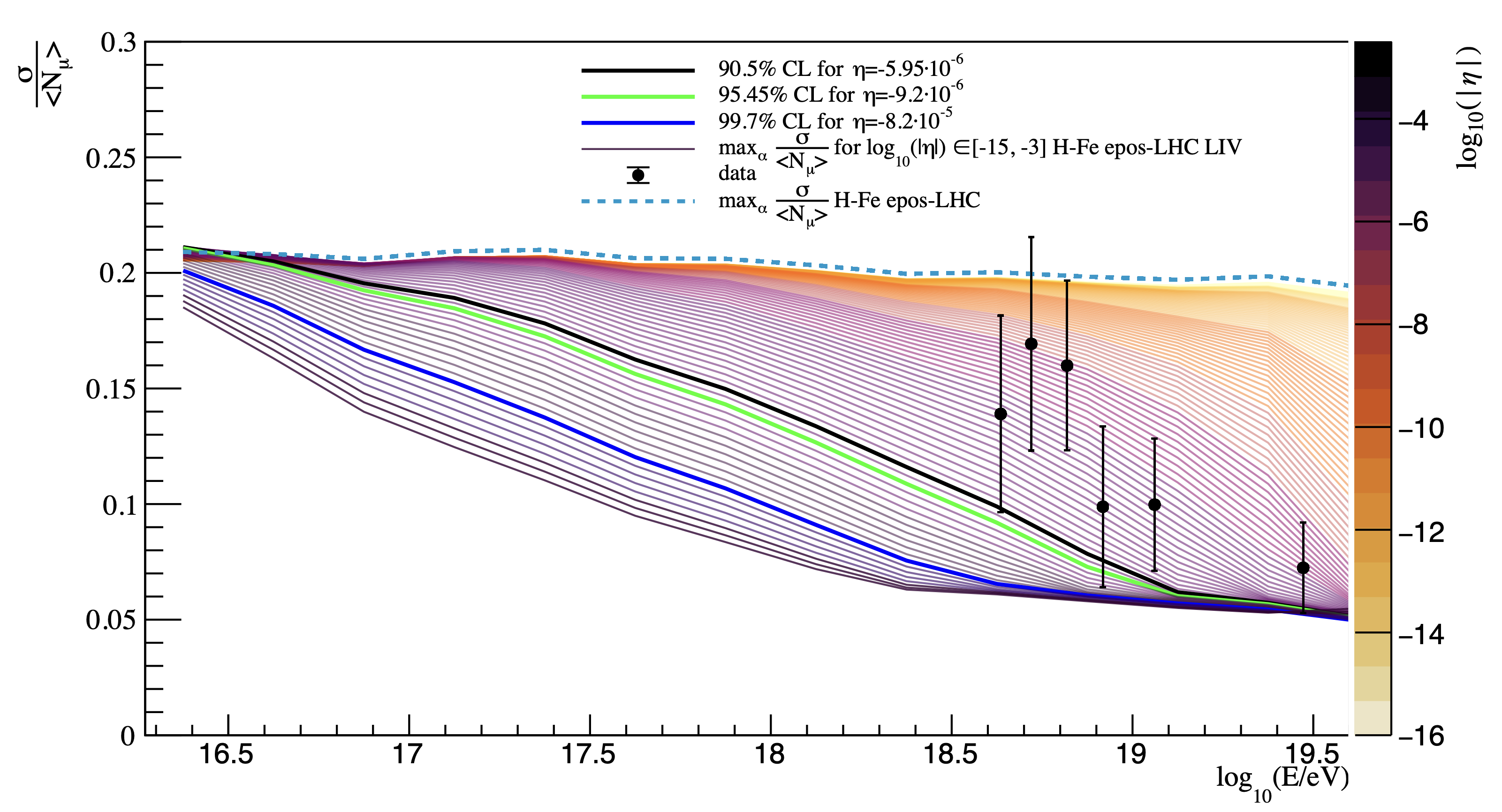}
    \caption{Relative fluctuations in the number of muons in EAS as a function of energy as measured at the Pierre Auger Observatory. Expectations from negative values of $\eta_{\pi^01}$ are shown as the colored curves. From~\cite{PierreAuger:2021mve}.}
    \label{fig:nmuliv}
\end{figure}

Lorentz-violating effects can be abrupt or gradual. Among abrupt ones is the suppression of the $\pi^0$ decay into two photons above some threshold that depends on negative values of $\eta_{\pi^0n}$. Conversely, the photon decay into another photon and a neutral pion becomes kinetically allowed. Such radical changes compared to the SM expectations can leave important signatures in the development of EAS. Indeed, neutral pion decay is in principle feeding in the first place the electromagnetic cascade. In the presence of Lorentz invariance violation, by contrast, neutral pions with energy above a certain threshold interact without decaying and therefore constitute an additional source of  hadronic sub-showers. Only after a few generations in the cascade, neutral pions with degraded energies behave in the standard way and initiate the electromagnetic component. Overall, Lorentz-invariance-violation effects lead to a reduced amount of energy deposited in the atmosphere that leads to an underestimation of the primary energy. Moreover, since the muon content of the shower correlates with the energy of the hadronic component, the relative fluctuations of the total muon number turns out to be smaller than in the standard case.  

At a fixed energy, the (negative) value of any $\eta_{\pi^0n}$ coefficient needs to be larger than $\eta_{\pi^0,n+1}$ by orders of magnitude to trigger decay suppression. Searching effects for thresholds larger than $10^6~$GeV, it is comfortable to assume $\eta_{\pi^00}$ to be negligible and to consider $\eta_{\pi^01}$ as the leading order. For different values of  $\eta_{\pi^01}$, simulations of EASs can be performed and used to search for unrealistic changes in the relative number of muons compared to data. To derive the most conservative bound on $\eta_{\pi^01}$, an optimal mixture of protons and iron nuclei is determined as a function of energy so as to maximize the relative fluctuations in the number of muons. Considering the parameter space $\eta_{\pi^01}\in[-10^{-3},-10^{-15}]$, the obtained relative fluctuations are shown as the colored thin curves in Fig.~\ref{fig:nmuliv}. Those too far away from the data points signal unrealistic underlying $\eta_{\pi^01}$ coefficients that can be excluded.  In this way, the following bound has been obtained with 90\% confidence level~\cite{PierreAuger:2021mve},
\begin{equation}
    \boxed{
    \eta_{\pi^01}>-6\times 10^{-6}.
    }
\end{equation}

\subsection{Gauge-invariant $CPT$ perturbation of QED?}

In the Standard Model Extension, one possible additional gauge-invariant term concerns the kinetic sector of the Maxwell field. It is built as a $CPT$-even perturbation of QED,
\begin{equation}
        \mathcal{L}=\mathcal{L}_\mathrm{QED}-\frac{1}{4}(k_F)_{\mu\nu\rho\sigma}F^{\mu\nu}F^{\rho\sigma},
\end{equation}
where $k_F$ encompasses 20 independent components including 19 degrees of freedom that lead to violation of Lorentz invariance. One of these components, $(k_F)^\lambda_{\mu\lambda\nu}$, causes an isotropic and non-birefringent modification of the photon propagation. This component is controlled by a single dimensionless parameter $\kappa$,
\begin{equation}
        (k_F)^\lambda_{\mu\lambda\nu}=\frac{\kappa}{2}[\mathrm{diag(3,1,1,1)}]_{\mu\nu},
\end{equation}
which relates the photon phase velocity to the maximal fermion velocity through \begin{equation}
c_\gamma=\left(\frac{1-\kappa}{1+\kappa}\right)^{1/2}c_\mathrm{f,max}.
\end{equation}
$\kappa$ also controls the changes in kinematical rules for some processes. For $\kappa<0$, neutral pion decay is suppressed above a certain threshold while photon decay is allowed, as in the case studied above. For $\kappa>0$ on the other hand, photon velocity gets smaller than the maximum attainable
velocity of fermions and vacuum Cherenkov radiation above a critical energy is possible.

With the possibility for high-energy electrons in EAS to radiate in vacuum, the descent in energy of electromagnetic sub-showers is faster than usually expected, leading to shallower values of the depth of shower maximum, $X_\mathrm{max}$, than those usually predicted for a given primary mass and energy. Eventually, too large values of $\kappa$ would lead to unrealistic shallow values of $X_\mathrm{max}$. On this basis, a bound can be set on $\kappa$ by comparing, for any combination of primaries,\footnote{Provided that above the threshold energy considered, the primary in question can make it to Earth without radiating.} the expected set of values in the plane $(\langle X_\mathrm{max}\rangle,\sigma(X_\mathrm{max}))$ to those measured at the Pierre Auger Observatory~\cite{Duenkel:2023nlk}. In this manner, the bound on positive $\kappa$ reads $\kappa<3\times10^{-20}$. 

On the other hand, photon decay and neutral pion stability can be studied in the same manner and a bound on negative $\kappa$ can be obtained~\cite{Duenkel:2021gkq}. Overall, the best constraints on $\kappa$ as derived from EAS are, within 98\% confidence level,
\begin{equation}\boxed{
    -6\times10^{-21}<\kappa<3\times10^{-20}.}
\end{equation}

\section{Ultra-high-energy gamma rays and neutrinos} \label{sec:UHEgnu}

\begin{wrapfigure}{L}{9 cm}
{\includegraphics[width=0.5\textwidth]{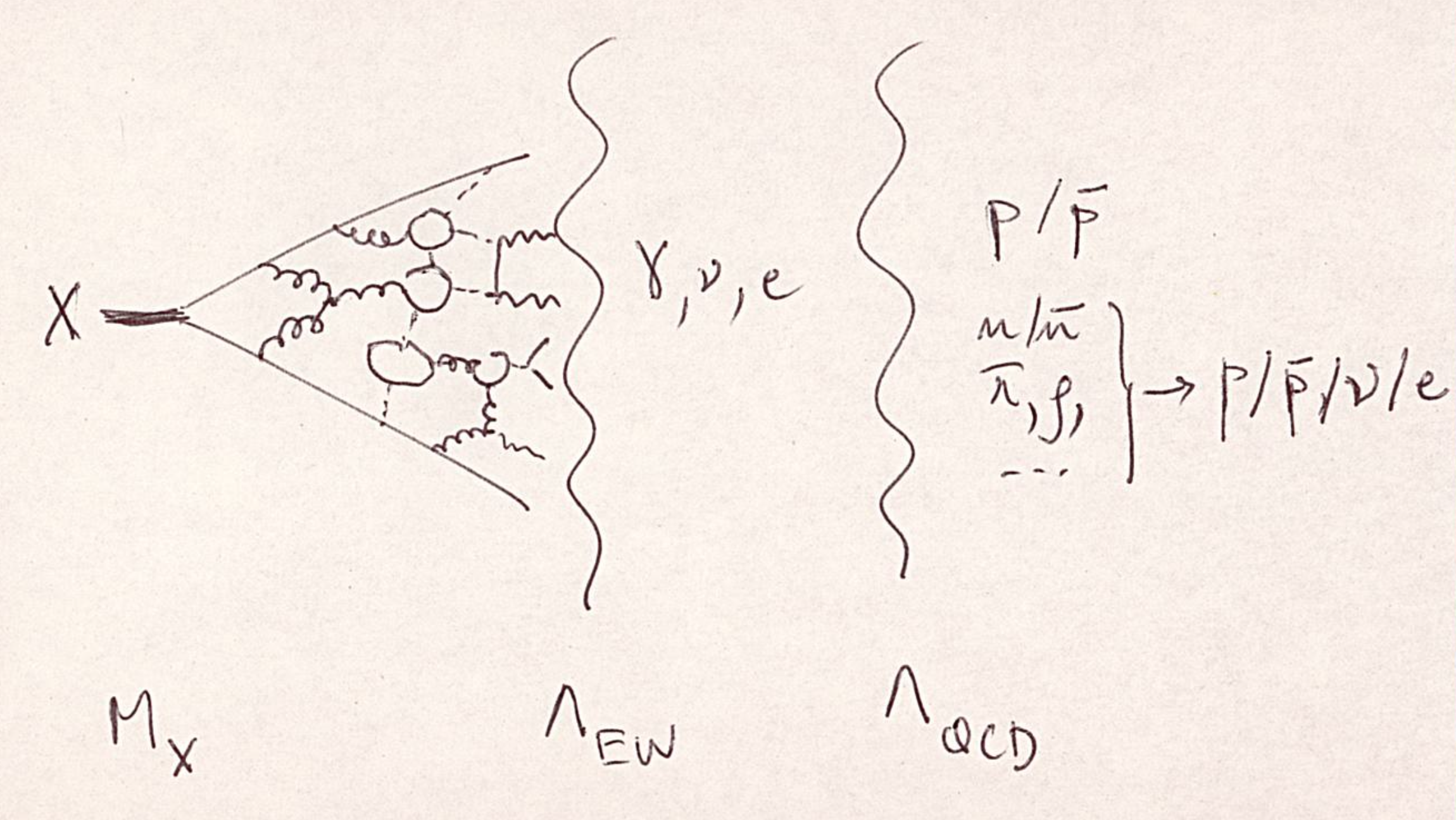}}
\caption{Cartoon illustration of fragmentation.}
\label{fig:fragmentation}
\end{wrapfigure}
Current limits on ultra-high-energy gamma rays and neutrinos fluxes are especially well suited to constrain high-scale BSM physics. Indeed, due to fragmentation effects for particles with mass much larger than the electroweak scale and \textit{a fortiori} the QCD transition scale, high and ultra-high energy particles, including nucleons, electrons, neutrinos and photons, are expected to emerge from the cascade subsequent to the decay of a superheavy particle or to the interaction of particles at high scale (Fig.~\ref{fig:fragmentation}). Seminal works in the QCD sector~\cite{Sarkar:2001se,Barbot:2002gt,Aloisio:2003xj} and in the electroweak one~\cite{Berezinsky:2002hq} have paved the way for the calculation of the fragmentation functions to derive the prompt flux of high-energy byproducts. 

For any particle of ultra-high energy $E$ of type $i$, the estimation of the diffuse flux (per steradian), $\phi_i(E,\mathbf{n})$, resulting from the prompt emission of decaying superheavy particles with lifetime $\tau_X$ is obtained by integrating the position-dependent emission rate per unit volume and unit energy along the ``lookback position''  in the direction $\mathbf{n}$, 
\begin{equation}
    \label{eqn:flux_i}
    \phi_i(E,\mathbf{n})=\frac{1}{4\pi}\int_0^\infty ds~q_i(E,\mathbf{x}(s,\mathbf{n})).
\end{equation}
For neutral particles, the lookback position reduces to the position along the line of sight, $\mathbf{x}(s,\mathbf{n})=\mathbf{x}_\odot+s\mathbf{n}$, with $\mathbf{x}_\odot$ the position of the Solar system in the Galaxy and $\mathbf{n}\equiv\mathbf{n}(\ell,b)$ a unit vector on the sphere pointing to the longitude $\ell$ and latitude $b$, in Galactic coordinates. The emission rate is shaped by the DM density $n_\text{DM}$, more conveniently expressed in terms of energy density $\rho_\text{DM}=M_X\,n_\text{DM}$, and by the differential decay width into the particle species $i$ as
\begin{equation}
    \label{eqn:q_i}
    q_i(E,\mathbf{x})=\frac{\rho_\text{DM}(\mathbf{x})}{M_X\tau_X}\frac{dN_i(E;M_X)}{dE}.
\end{equation}
The ingredients are thus well separated in terms of particle-physics inputs, which regulate the spectra of secondaries from the decaying particle, and astrophysical factor. 

Secondary electrons with energy reaching possibly $10^{13}~$GeV may also be relevant~\cite{Munbodh:2024ast,Deligny:2024fyx}, although they get rapidly attenuated in the Galaxy and do not reach the Earth. Secondary fluxes of ultra-high energy gamma-rays from inverse Compton-scattering on low-energy photons from cosmic microwave, infra-red or star-light backgrounds are negligible compared to prompt fluxes. However, secondary fluxes from synchrotron emission can be significant for $M_X$ large enough. The emission rate entering into Eqn.~\ref{eqn:flux_i} can be customarily written as
\begin{equation}
    \label{eqn:q_gamma_syn}
    q_\gamma(E,\mathbf{x})=\frac{1}{E}\int_{m_\mathrm{e}}^{M_X/2}d E_\mathrm{e}\frac{d n(E_\mathrm{e},\mathbf{x})}{d E_\mathrm{e}}\frac{d P_\mathrm{syn}(E_\mathrm{e},E,\mathbf{x})}{d E}.
\end{equation}
Here, $dP_\mathrm{syn}(E_\mathrm{e},E,\mathbf{x})/d E$ is the differential synchrotron power  of electrons with energy $E_\mathrm{e}$ (and mass $m_\mathrm{e}$) emitting photons in the band between $E$ and $E+dE$ (see, e.g., ~\cite{Leite:2016lsv}), and $dn/dE_\mathrm{e}$ is the differential density of electrons. The latter can be approximated by as
\begin{equation}
    \label{eqn:ne}
     \frac{d n(E_\mathrm{e},\mathbf{x})}{d E_\mathrm{e}}=\frac{\rho_\text{DM}(\mathbf{x})}{M_X\tau_Xb(E_\mathrm{e},\mathbf{x})}Y_\mathrm{e}(E_\mathrm{e}),
\end{equation}
with $Y_\mathrm{e}(E_\mathrm{e})=\int_{E_\mathrm{e}}^{M_X/2}d E'_\mathrm{e}~dN_\mathrm{e}(E'_\mathrm{e};M_X)/dE'_\mathrm{e}$ the yield of electrons with energy larger than $E_\mathrm{e}$, and $b(E_\mathrm{e},\mathbf{x})$ the energy-loss rate due to synchrotron emission and inverse Compton scattering (see, e.g.,~\cite{Blumenthal:1970gc}). For electrons with energy above $10^{13}~$GeV, the flux of gamma rays from synchrotron radiation turns out to dominate over that from prompt emission.

\section{Superheavy dark matter} \label{sec:SHDM}

The mass $M_X$ of dark matter (DM) particles that do not undergo weak interactions is not generated by the Higgs vacuum expectation value and is therefore not tied to the electroweak scale. In other words, $M_X$ is unconstrained.  One intriguing energy range for BSM physics, as an intermediate step below the Grand Unification scale, lies between $10^{10}$ and $10^{13}~$GeV. This range encompasses the mass of the inflaton, indirectly inferred to lie around $(1$--$3)\times 10^{13}~$GeV, the mass of the right-handed neutrinos  within the vanilla seesaw mechanism, and the instability scale of the Standard Model inferred within $10^{10}$-to-$10^{12}$\,GeV from LHC data.  The mass spectrum of the dark sector could also reflect this high-energy scale, and  various mechanisms in the framework of inflationary cosmology are capable of producing superheavy DM particles. A minimal and unavoidable process is gravitational production, which can take place during the inflationary era~\cite{Chung:1998zb,Kuzmin:1999zk} and/or the reheating one~\cite{Garny:2015sjg,Mambrini:2021zpp}. 

Superheavy particles must be extremely long-lived, $\tau_X\gtrsim 10^{22}~$yr, to be responsible for the relic density of DM observed today. Only a handful theoretical constructions can meet these constraints without resorting to fine-tuning: sterile neutrinos being themselves DM and feebly coupled~\cite{Uehara:2001wd,Feldstein:2013kka,Dev:2016qbd} or with slight mass-mixing in the sterile neutrino sector~\cite{Rott:2014kfa}, particles coupled with sterile neutrinos alone~\cite{Dudas:2020sbq,PierreAuger:2023vql}, or non-perturbative effects suppressing the dark coupling constant and selecting large-multiplicity final states from instantonic rules~\cite{Kuzmin:1997jua,PierreAuger:2022jyk}. Another possibility highlighted in this review relies on spin $3/2$ particles in the context of supersymmetry broken at high scale. 

If the gravitino is the lightest supersymmetric particle, it may be a natural candidate to DM. It is well established, however, that its production from the decay of the next-to-lightest supersymmetric particle may far exceed the relic abundance observed today.  One way to sidestep this problem is to invoke that the mass spectrum of supersymmetric partners is such that only the gravitino could be created in the thermal bath after inflation~\cite{Dudas:2017rpa}. This implies a mass scale above the reheating temperature $T_\mathrm{rh}$ to prevent creation from thermal processes, and above the mass of the inflaton to prevent creation from inflaton decay. Together, these conditions allow the gravitino mass $M_{3/2}$ to be constrained such that
\begin{equation}\label{eqn:M32}
    M_{3/2}\geq\frac{M_\phi^2}{\sqrt{3}M_\mathrm{P}}\simeq 2\times 10^8~\mathrm{GeV}
\end{equation}
for an inflaton mass $M_\phi\simeq 3\times 10^{13}$~GeV~\cite{Dudas:2017rpa}.

\begin{wrapfigure}{L}{8 cm}
{\includegraphics[width=0.5\textwidth]{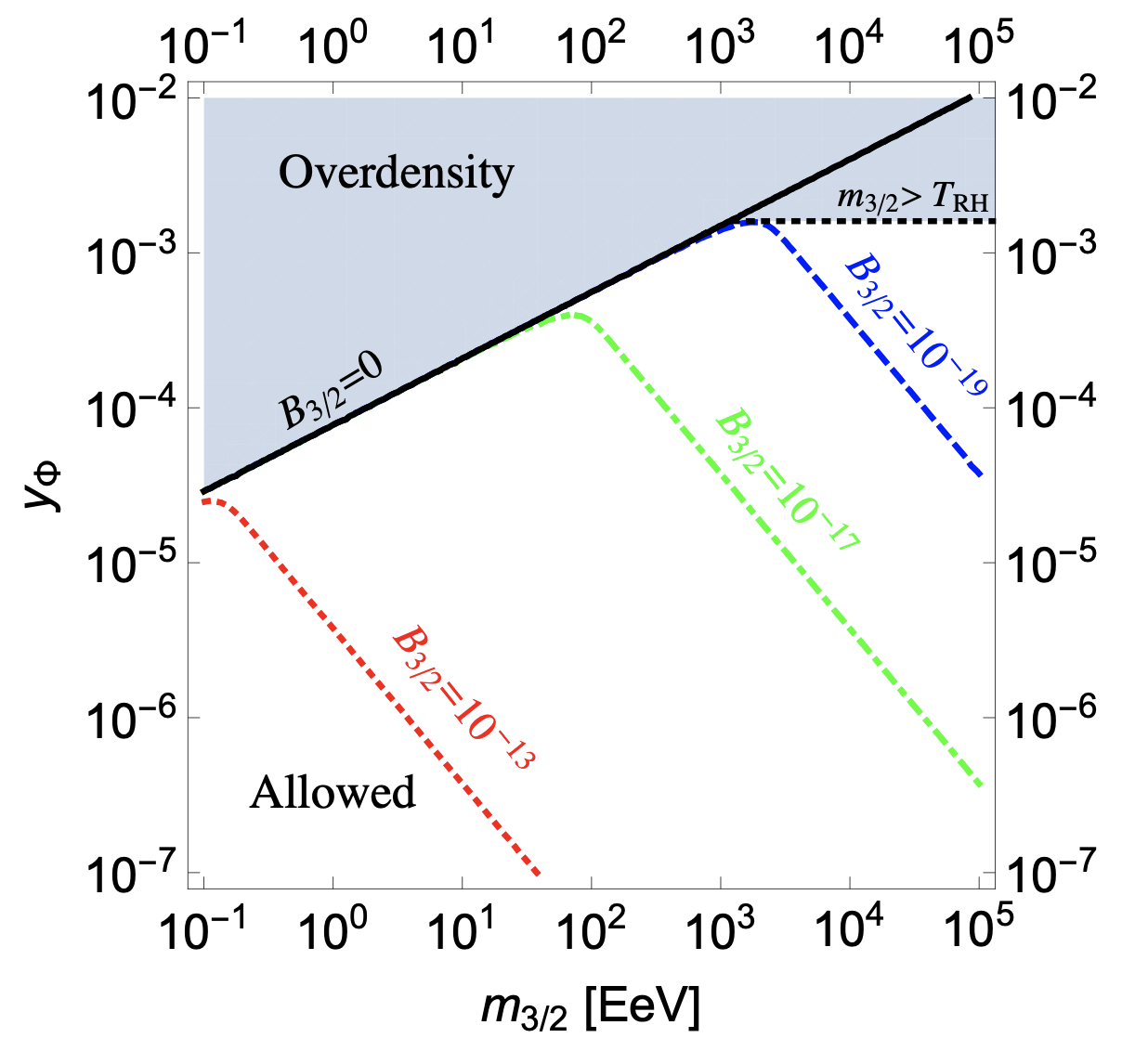}}
\caption{Viable values of Yukawa-like coupling of inflaton to matter and of gravitino mass to match the relic density of DM, for different values of branching ratio of inflaton decay to gravitinos. From~\cite{Dudas:2017rpa}.}
\label{fig:gravitino}
\end{wrapfigure}
In this scenario, the gravitino could be created in two ways. The first one is through thermal production in the s-channel via the gluon+gluon $\rightarrow$ gravitino+gravitino process, which is the only one allowed kinematically. Then the relic density scales as
\begin{equation}
    \frac{\Omega_{3/2}h^2}{0.12}=\left(\frac{M_{3/2}}{10^8~\mathrm{GeV}}\right)^{-3}\left(\frac{T_\mathrm{rh}}{10^{10}~\mathrm{GeV}}\right)^7,
\end{equation}
which requires a high reheating temperature. This scaling can be recast in terms of the ``Yukawa''-like coupling $y_\phi$ of the inflaton field to the thermal bath instead of the reheating temperature. The viable range of parameters in the plane $(y_\phi,M_{3/2})$ is shown as the black curve labeled as $B_{3/2}=0$ in Fig.~\ref{fig:gravitino}. The linear increase in $y_\phi$ with increasing gravitino mass is necessary to counterbalance the weakening of the eﬀective coupling scaling as $1/M^2_{3/2}$. Note in addition the upper bound shown as the dotted black line (labeled as $M_{3/2}>T_\mathrm{rh}$), stemming from the condition that the gravitino mass must be less than the reheating temperature by construction.  Alternatively, gravitinos coupled to moduli fields can also be produced by direct decay of the inflaton. The condition~(\ref{eqn:M32}) then translates into a bound for the branching ratio $B_{3/2}$ of the inflaton decay into gravitinos,
\begin{equation}
    B_{3/2}\leq 1.9\times10^{-18}\left(\frac{1}{y_\phi}\right)\left(\frac{M_{3/2}}{10^8~\mathrm{GeV}}\right)^{-1}.
\end{equation}
The viable parameter space is also shown in Fig.~\ref{fig:gravitino} as the color-coded dotted curves. For a given value of $B_{3/2}$, the Yukawa coupling is required to be smaller than in the  $B_{3/2}=0$ case for masses above the turning points, signaling that the direct production from inflaton decay  dominates over the thermal production. 

If the $R$-parity is exactly conserved, the lightest supersymmetric particle, namely the gravitino in this setup, is stable and there are very few detectable signatures of the model. A tiny violation of the $R$ parity, however, might be sufficient to render the gravitino metastable and lead to observational signatures in ultra-high-energy gamma rays and neutrinos. Stringent limits on $R$-parity-violation couplings arise from preserving baryon asymmetry in the early universe by requiring that interactions violating the number of baryons and leptons remain out of equilibrium. However, for supersymmetric partners never in the thermal bath to mediate interactions that would wash out the baryon asymmetry, these limits are quite relaxed. Considering a single $R$-parity violating interaction with a bilinear term in the superpotential such as $\mu'LH_u$ gives rise to dominant decay channels to $Z\nu$, $We/W\mu/W\tau$ and $h\nu$ final states~\cite{Dudas:2018npp,Allahverdi:2023nov}. By calculating the fluxes of secondary gamma rays and neutrinos emitted in each decay channel as a function of $M_{3/2}$, upper limits on $\mu'$ can be obtained by requiring these fluxes not to overshoot the limits. Roughly, the constraints can be summarized as
\begin{equation}\boxed{
    \mu'\lesssim10^{-5}\left(\frac{M_{3/2}}{10^8~\mathrm{GeV}}\right)^{-2}~\mathrm{GeV}.}
\end{equation}
The limits are to be compared to those obtained in the context of weak-scale supersymmetry, namely $\mu'<20$~keV from the preservation of the baryon asymmetry.

\section{Cosmic strings} \label{sec:CS}

Phase transitions may have occurred in the early universe provided that the reheating temperature was high enough. Many theoretical setups predict the existence of $U(1)$ symmetries at high scale for which phase transitions  would have led to the formation of cosmic strings, which are regions of space-time that remain in a symmetry unbroken phase due to boundary conditions that topologically restrict their decay. For this to happen, the $U(1)$ symmetry must have completely disappeared after the transition.\footnote{This is why the electroweak transition does not lead to the formation of cosmic strings due to the particular encapsulation of electromagnetism in $SU(2)_L\times U(1)_Y$.} 

\begin{wrapfigure}{L}{9 cm}
{\includegraphics[width=0.5\textwidth]{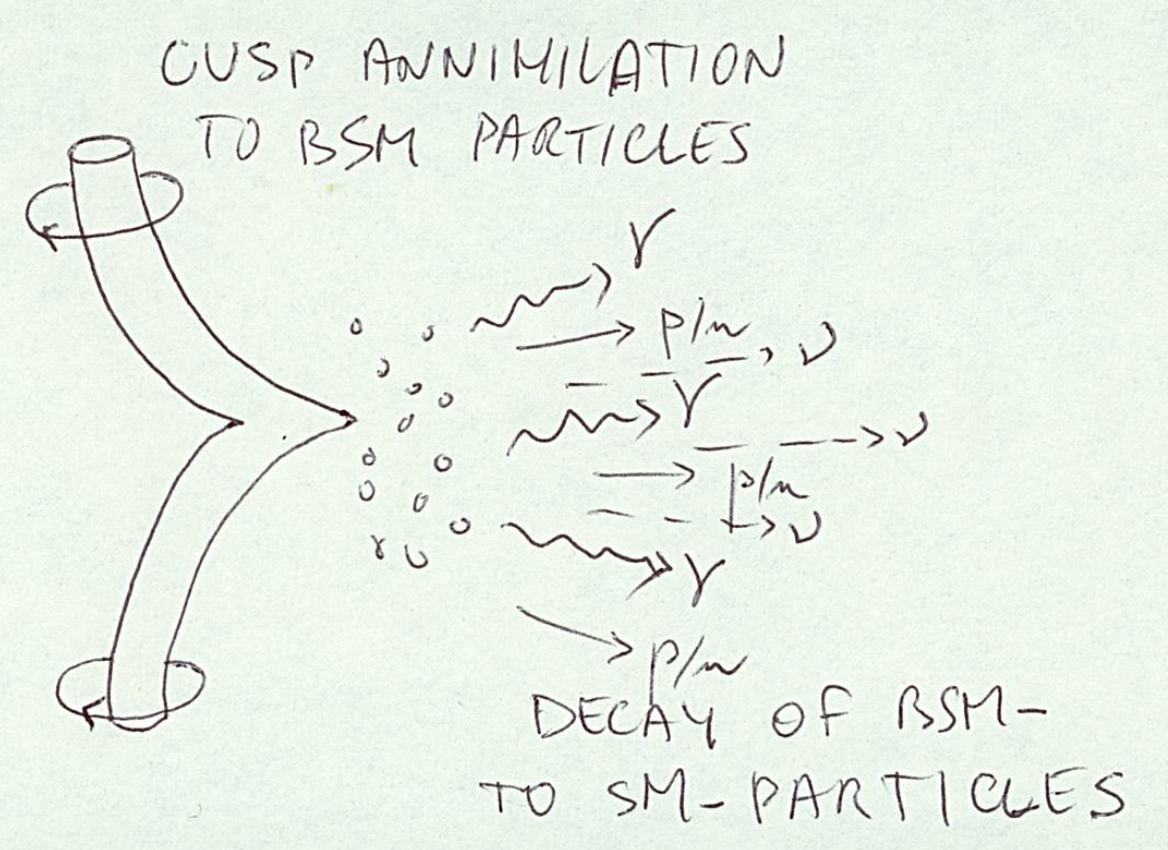}}
\caption{Cartoon illustration of cusp annihilation.}
\label{fig:cusp}
\end{wrapfigure}
Under certain circumstances, yet, the energy stored in the unbroken vacuum phase can be liberated in the form of high-scale quanta of the fields. This is the case in particular when the dynamics of the strings leads to ``cusps'', which are short segments with velocity momentarily very close to $c$. Emission from cusps results in particles with ultra-high energies due to the large Lorentz factors at play (see Fig.~\ref{fig:cusp}). Note that the emitted cosmic rays are charged particles in the case of superconducting strings or Higgs-decay byproducts in the case of the development of a bosonic condensate around the string core. Alternatively, as explored in~\cite{Berezinsky:2011cp}, cusps could emit moduli, which are relatively light, weakly coupled scalar fields, predicted in supersymmetric and string theories. Subsequently, a modulus decays into two gluons, which initiate a quark-gluon cascade that produces hadrons, mostly pions, and eventually protons, photons, and neutrinos. 

\begin{wrapfigure}{L}{9 cm}
{\includegraphics[width=0.5\textwidth]{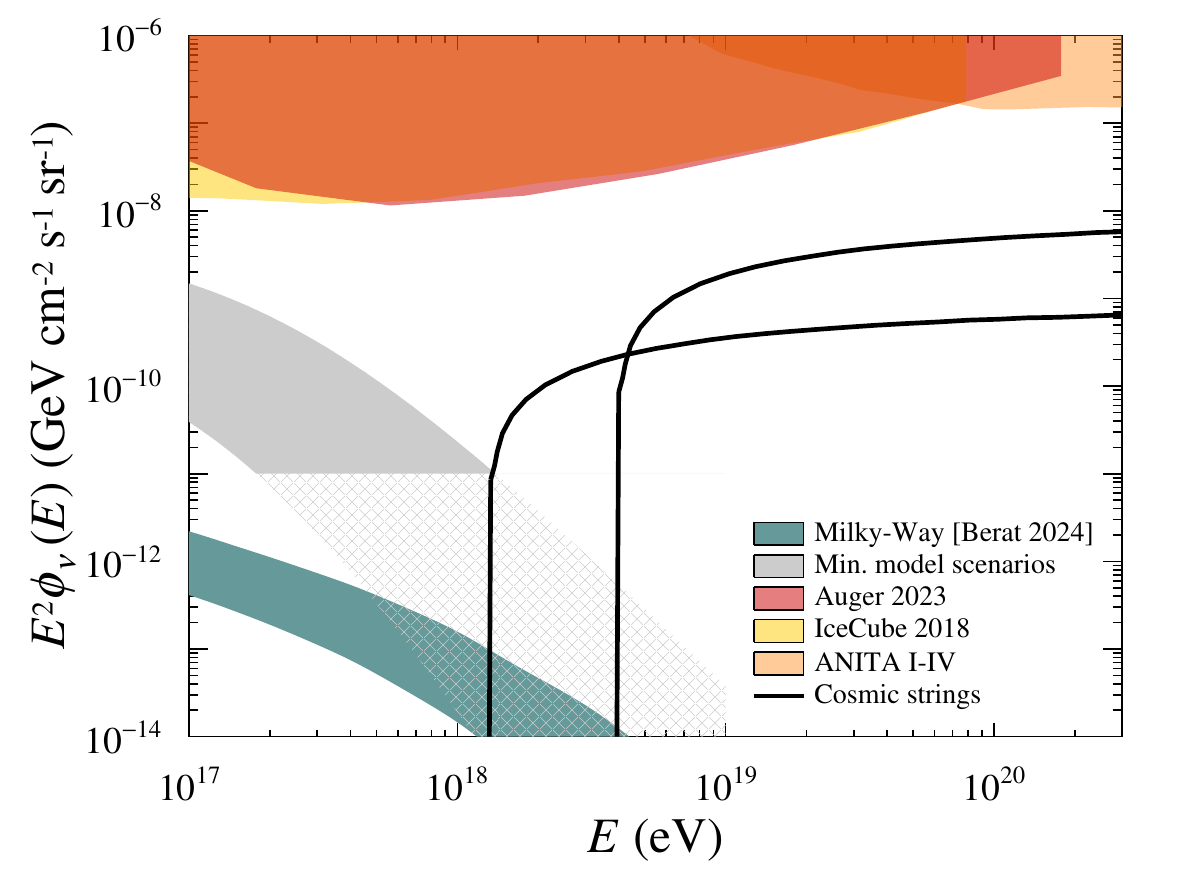}}
\caption{Energy flux of neutrinos (single flavor) expected from cosmic-ray interactions in the milky way (``Milky Way'') or from various source environments that fit with the minimal model explaining the Auger data above $10^{9.7}~$GeV (``Min. model scenarios'')~\cite{Berat:2024rvf}. Upper limits from Auger, IceCube and ANITA are reported on top. Also shown are energy fluxes expectations from decay of cosmic strings (see text).}
\label{fig:nulimits}
\end{wrapfigure}
The horizon of neutrinos is characterized by a redshift scaling as $z_\nu\simeq 250~(E/10^{11}~\mathrm{GeV})^{-2/5}$, which is particularly relevant in this context as the redshift $z_\star$ at which the emission is dominant corresponds to the epoch of transition between gravitation-dominated and modulus-dominated radiation, $z_\star\simeq 400~m_5^{2/3}\alpha_7^{-8/3}\mu_{-20}^{2/3}$, with $m_5$ the modulus mass in units of $10^5~$GeV, $\alpha_7$ the coupling between the modulus and the stress energy tensor of the strings in units of $10^7$, and $\mu_{-20}$ the string tension in units of $10^{-20}G\mu$. The energy spectrum of the neutrinos emitted subsequently to the decay of moduli into gluons can be approximated as an $E^{-2}$ one with a low-energy cutoff due to the suppression of soft gluon emission in the parton cascade. Once boosted by a relativistic factor $\gamma$, the concentration of high-energy particles in cones whose opening angles $1/\gamma$ point in the direction of the boost results in the appearance of a low-energy end of the spectrum. For $z_\nu=200$, the low-energy cutoff scales as $\sim 4\times 10^9m_5^{1/2}\mu_{-20}^{1/2}$~GeV, which nicely ranges within the sensitivity of, e.g., the Pierre Auger Observatory and brings clear signatures for this model. 

With a model of the rate of cusp bursts $d\dot{N}_b$ from simulations and a model for the number of moduli $dN^b_X(k)$ emitted in a single burst with momenta $k$ in the interval $dk$, the flux of neutrinos on Earth can be calculated as~\cite{Berezinsky:2011cp}
\begin{equation}
    J_\nu(E,z)=\frac{1}{(4\pi)^2}\int \frac{dV(z)}{(1+z)r^2(z)}d\dot{N}_b~dN^b_X(k)~\frac{dN_\nu(E,z,k)}{dE},
\end{equation}
with $dV(z)$ the proper volume and $dN_\nu/dE$ the spectrum of neutrinos from decay of a
modulus with momentum $k$. Several neutrino fluxes corresponding to different choices of parameters and conforming to the limits obtained with IceCube, the Pierre Auger Observatory and ANITA are illustrated in Fig.~\ref{fig:nulimits}. Generic estimates of cosmogenic fluxes based on a minimal model of UHECR production to explain the mass-discriminated energy spectra observed on Earth above $5{\times}10^{18}~$eV are shown as the gray band~\cite{Berat:2024rvf}. One striking feature of the fluxes from cosmic-string decay is the sharp rise at energies above the suppression ones for cosmogenic fluxes preceding an $\sim E^{-2}$ recovery. Finally, in addition to probing BSM physics at scale of $\sim 10^5$~GeV thanks to the boost factors, and because $J_\nu$ is approximately proportional to $(G\mu)^3$, the detection of neutrino fluxes featuring the specific signatures highlighted above would make it possible to probe cosmic string tensions as low as 
\begin{equation} \boxed{
  G\mu\lesssim10^{-20},}  
\end{equation}
which is to be compared to the best-limits currently obtained with LIGO/VIRGO of order of $G\mu=10^{-15}$.

\section{BSM with upward-going showers} \label{sec:up}

In the SM, the neutrino-nucleon scattering cross-section increases with the energy of the incoming neutrino. Consequently, ultra-high-energy neutrinos may only propagate through the Earth for relatively short distances of the order of $\mathcal{O}(100)$~km. Motivated by two ``anomalous'' radio pulses observed with the ANITA instrument compatible with EASs developing in the upward direction and inconsistent with SM expectations~\cite{ANITA:2016vrp}, a dedicated search for upward-going EASs at zenith angles exceeding 110$^\circ$ and energies $E >10^8~$GeV has been performed using the Fluorescence Detector of the Pierre Auger Observatory~\cite{PierreAuger:2023elf}. One event candidate has been found, consistent with an expected background of $0.27 \pm 0.12$ events from mis-reconstructed UHECR showers. Upper bounds on the integral flux of $(7.2 \pm 0.2) \times 10^{-21}~$cm$^{-2}$~sr$^{-1}$~yr$^{-1}$ and $(3.6 \pm 0.2) \times10^{-20}~$cm$^{-2}$~sr$^{-1}$~yr$^{-1}$ have been derived for $E^{-1}$ and  $E^{-2}$ spectra, respectively.  These limits are stringent enough to exclude a physics origin of the two anomalous ANITA events.

\begin{wrapfigure}{L}{9 cm}
{\includegraphics[width=0.5\textwidth]{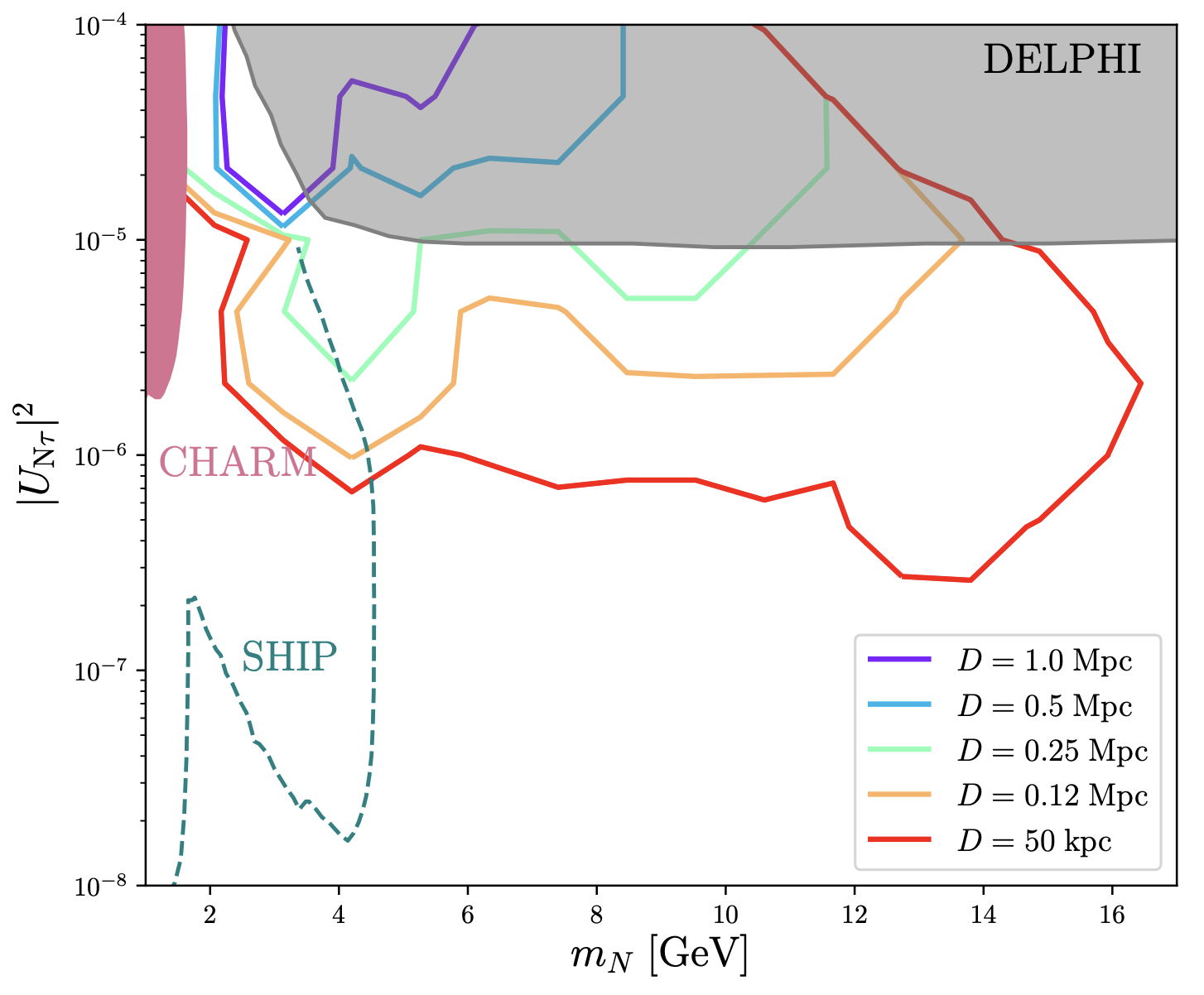}}
\caption{Sensitivity at 99.7\% confidence level of POEMMA to sterile neutrinos for an emergence angle of 60$^\circ$ and considering a Gamma Ray Burst
similar to GRB221009A located at a distance $D$. From~\cite{Heighton:2023qpg}.}
\label{fig:constraints-N}
\end{wrapfigure}
The claim from ANITA has attracted much attention from particle physics theorists, who have sought to establish BSM scenarios capable of matching anomalous events. Building on these ideas, sensitivity to upward-going EAS has proved to be a powerful tool for probing scenarios otherwise difficult to explore with colliders. One compelling scenario is that of an upscattering of $\tau$ neutrino into a sterile one that would traverse the Earth unattenuated before decaying into a $\tau$ neutrino and/or a $\tau$ close enough to the Earth surface to give rise to a classical Earth-skimming event. Provided an exceptional transient source such as GRB221009A to produce neutrino extending to ultra-high energies, the upward-going channel could enable the detection of deca-GeV sterile neutrinos~\cite{Heighton:2023qpg}. The BSM setup consists here in the minimal extension of SM Lagrangian with one $N_\mathrm{R}$ degree of freedom coupling only to the $\tau$ flavor, 
\begin{equation}
\mathcal{L}_\mathrm{BSM}=\mathcal{L}_\mathrm{SM}+\frac{m_N}{2}\overline{N}^\mathrm{c}_\mathrm{R}N_\mathrm{R}-\frac{g}{\sqrt{2}}\sin\theta_\mathrm{mix}W_\mu^+\overline{N}^\mathrm{c}_\mathrm{R}\gamma^\mu \tau_\mathrm{L}-\frac{g}{2\cos\theta_\mathrm{mix}}\sin\theta_\mathrm{mix}Z_\mu\overline{N}^\mathrm{c}_\mathrm{R}\gamma^\mu \nu_\tau+\mathrm{h.c.},    
\end{equation}
with $\theta_\mathrm{mix}$ controlling the mixture between active and sterile degrees of freedom in the (Majorana) neutrino mass eigenstates.  In this way, right-handed  degrees of freedom inherit from SM couplings to matter but reduced by $\theta_\mathrm{mix}$. Using the incoming $\tau$ neutrino fluence from a Gamma Ray Burst similar to GRB221009A located at a distance $D$, and modeling the exposure of POEMMA to upward-going showers, the number of events expected at large emergence angle and within $10^9$ and $10^{11}~$GeV can be calculated under both the SM and the BSM assumptions. Requiring a rejection of the BSM hypothesis with a 99.7\% confidence level, an interesting region in the plane $(\theta_\mathrm{mix}^2,m_N)$ can be probed as shown in Fig.~\ref{fig:constraints-N}. The sensitivity that would be achieved this way is competitive with existing constraints and may provide complementary probes compared to future long-lived particle search experiments.

\section{And next} \label{sec:future}

The near future of UHECR science will be the determination with the upgraded Pierre Auger Observatory of the mass composition at ultra-high energy. The gradual increase with energy from CNO to Si and possibly Fe group elements would confirm the minimal setup for UHECRs explained by a single population of extragalactic sources above the ankle energy. In such a case, the Observatory would have fulfilled its initial mission of providing answers to the origin and nature of UHECRs, and its data would be relevant for studying the origin, dynamics and energetics of relativistic jets and compact objets, or for the astrophysics of magnetized plasmas in motion or of strong gravity environments. 

On the other hand, the presence of a sub-dominant component of protons would re-open the good-old issue of acceleration at such high rigidities. A natural alternative to acceleration would be BSM physics, and the various signatures in the sector of gamma rays and neutrinos would be instrumental in deciphering which BSM physics is at play. Based on concrete particle-physics setups, a few emblematic possibilities have been addressed in this review. UHECRs remain indeed the ultimate laboratory of high-energy physics.

\bibliographystyle{JHEP}
\bibliography{biblio}

\end{document}